\documentclass[twocolumn,showpacs,prb,aps]{revtex4}

\usepackage{graphicx}
\usepackage{dcolumn}
\usepackage{bm}
\usepackage{color}
\usepackage{subfigure}

\begin{document}

\title{ Terahertz wave generation from hyper-Raman lines in two-level quantum systems driven by two-color lasers }

 \author{Wei Zhang {\footnote
{Author to whom any correspondence should be addressed, {Email:
zhang$\_$wei@iapcm.ac.cn}}}}
 \affiliation{Institute of Applied
Physics and Computational Mathematics, P. O. Box 8009(28), Beijing
100088, China}
\author{Shi-Fang Guo}\affiliation{Institute of
Applied Physics and Computational Mathematics, P. O. Box 8009(28),
Beijing 100088, China}
\author{Su-Qing Duan}
\affiliation{Institute of Applied Physics and Computational
Mathematics, P. O. Box 8009(28), Beijing 100088, China}
\author{Xian-Geng Zhao}
\affiliation{Institute of Applied Physics and Computational
Mathematics, P. O. Box 8009(28), Beijing 100088, China}

\date{\today}
\begin{abstract}
Based on spatial-temporal symmetry breaking mechanism, we propose a
novel scheme for terahertz (THz) wave generation from hyper-Raman
lines associated with the 0th harmonic (a particular even harmonic)
in a two-level quantum system driven by two-color laser fields. With
the help of analysis of quasi-energy, the frequency of THz wave  can
be tuned by changing the field amplitude of the driving laser. By
optimizing the parameters of the laser fields, we are able to
 obtain arbitrary
frequency radiation in the THz regime with appreciable strength (as
strong as the typical harmonics). Our proposal can be realized in
experiment in view of the recent experimental progress of
even-harmonics generation by two-color laser fields.
\end{abstract}
\maketitle

{\bf Introduction} Terahertz (THz) radiation, electromagnetic
radiation with typical frequency from 0.1 THz to 10 THz-lying in the
spectrum gap between the infrared and microwaves, has varieties of
applications in information and communication technology, biology
and medical sciences, homeland security, and global environmental
monitoring etc. \cite{1Tonouchi2007}. The generation of THz wave is
a challenge problem due to the lack of appropriate materials with
 small bandgaps in the usual optical approach. Great
effort has been made to the design of THz sources
\cite{17Otsuji2006,18Sekine2005,19Orihashi2005,20Crowe2005,
16Ito2005,8Kawase2001,fwmix,10Faist1994,11Kohler2002,14williams2007,hu2004,9Belkin2007,
21Williams2002,22Bergner2005}. THz wave can be obtained by both
electronic
and optical methods.
The  uni-travelling-carrier photodiode \cite{16Ito2005} and the
quantum cascade
laser\cite{10Faist1994,11Kohler2002,14williams2007,hu2004,9Belkin2007}
  are two examples. The electronic approach is limited in the low
frequency end of the THz regime, and the optical approach usually
focuses on the cases of small energy gap. Technological innovation
in photonics and nanotechnology has provided us more new ways for
THz radiation generation. Recently an interesting approach based on
semiconductor nanostructure driven by acoustic wave was suggested
\cite{22a Ahn2007}, and electrically pumped photonic-crystal THz
laser was developed \cite{22b Chassagneux2009}. THz wave generation
by up conversion method through high-order harmonics generation(HHG)
in semiconductor quantum dot(QD) was also suggested in our previous
work \cite{3zhang2009PRB}.

One challenge of THz wave generation in optical approach in usual
system is to generate low frequency radiation from a quantum system
with a large energy gap. Usually, in the emission spectrum of a
quantum system, there are harmonics, as well as the associated
hyper-Raman lines. Hyper-Raman lines are caused by the transitions
between the dressed bound states. Nth hyper-Raman lines are defined
as those associated with the nth harmonics (they appear with the
corresponding harmonics simultaneously due to the same symmetry
reason, and locate near $n\omega_0$, where $\omega_0$ is the
fundamental frequency of the incident laser) \cite{23Millack}. Here
is one observation that the shift of the hyper-Raman line to the
associated harmonic could be small and could be tuned by the
external field, even though the fundamental frequency is large. Then
one may use the 0th hyper-Raman line (that associated with the 0th
harmonic) to generate the low-frequency (THz) wave. There have been
some studies on the harmonics and hyper-Raman lines. In many cases,
there are only odd harmonics due to the particular symmetry and the
associated hyper-Raman line is of high frequency and/or weak
intensity. Therefore one has to solve these two problems to generate
THz wave by using the hyper-Raman lines in the nonlinear optical
processes.

In this article, we adopt an effective optical method to obtain THz
wave in typical two-level quantum systems, which solves the two
problems mentioned above. Based on  the generalized spatial-temporal
symmetry principle we developed recently \cite{30sf2011NJP},  we
propose a novel mechanism to obtain the 0th hyper-Raman lines in THz
regime by introducing additional laser with frequency $2k\omega_0$
($k=1,2,3,...$). Using analysis of quasienergy and optimization of
laser fields,
 we are able to obtain considerably intense THz radiation with
 desired frequency. With the help of our optimization method, it is
 quite likely that our proposal can be realized in
 experiment considering the recent experimental progress of even-harmonics
 generation by two-color laser.

{\bf Theoretical formulism} Our two-level quantum system driven by
laser fields [see the schematic diagram Fig. 1(a)] is described by
the Hamiltonian
\begin{eqnarray}
H=\sum_i^{2} E_i |i\rangle \langle i |+G(t) (|1\rangle \langle 2|
+|2\rangle \langle 1|), \label{H}
\end{eqnarray}
where $G(t)= F(t)\textbf{e}\cdot\bm{\mu_{12}}$, is the Rabi
frequency caused by periodic laser field
$\textbf{E}=F(t)\textbf{e}$, $F(t)=F(t+T)$, $\textbf{e}$ a unit
vector. ${\bm \mu_{12}} =\langle 1 |e \textbf{r} |2\rangle$ is the
dipole between state $|1\rangle$ and state $|2\rangle$. The energy
spacing between the two states [with energies $E_i(i=1,2)$] is set
as $\triangle E =E_1-E_2$.

The dynamics of our system is described by the equation of motion of
the density matrix \cite{30Narducci1990PRA}
\begin{eqnarray}
\frac{\partial \rho}{\partial
t}=-\frac{i}{\hbar}[H,\rho]-\Gamma\cdot\rho \label{D1},
\end{eqnarray}
where $H$ is the Hamiltonian, the last term describes possible
dissipative effects (such as spontaneous phonon  emission) and we
set $\hbar=1$ in the following. We numerically calculate the photon
emission spectra of the two-level quantum systems by solving the
density matrix in Eq.(\ref{D1}) through Runge-Kutta method with the
time step of $0.002/\omega_0$, total steps of 1200000 and the
electron initially setting at the lower level. The average dipole
can be calculated as $\textbf{D}(t)=\sum_{ij} {\bm \mu_{ij}}
\rho_{ij}(t)$.   We use Fourier transformation to obtain the
emission spectrum $S(\nu)=|\int dt \textbf{exp}(-i\nu
t)\textbf{D}(t)|^2$. In the calculation we set $\hbar\omega_0$
($\omega_0$ the driving field frequency) as the unit of energy. We
choose $\omega_0=100THz$ as an example, yet in general one can also
use other frequency (much larger than THz) laser as will be
discussed later. The two-level quantum systems can be realized in
many systems, such as atoms, molecules, and semiconductor quantum
dots. Here we use the typical parameters of a quantum dot: the
energy spacing $\Delta E=10\hbar\omega_0$, the dipole moment
$\mu_{12}=0.5~ e\cdot nm$, phonon emission coefficient
$\Gamma_{ij}=3.4\times10^{-3}\omega_0$.

 We explore the emission spectra
using above numerical approach and the theory on the selection rule
in high-order harmonic generation based on the generalized
spatial-temporal symmetry. As described in our previous
work\cite{30sf2011NJP}, for a quantum system described by
Hamiltonian (1), if there exists one symmetric operation Q, which is
the time shift $\theta: t \rightarrow t+T/2$ ($T=2\pi/\omega_0$)
combined with another operation $\Omega$ in spatial/spectrum domain,
i.e. $Q=\Omega\cdot \theta\ $, such that the initial condition and
the Hamiltonian (up to a sign) are invariant, and the dipole
operator $\hat{P}$ has a definite parity, then the emission spectrum
contains no odd/even component if operator $\hat{P}$ is even/odd.
For our two-level quantum system driven by a monochromatic laser,
i.e. $E(t)=F_1cos(\omega_0t)$, there exists one symmetric operation
$Q_1$, which is the time shift $\theta$
combined with spatial operation $\Omega_1$: $c_1 \rightarrow -c_1$
(here and in the following $c_j, j=1,2$, refers to the annihilation
operator for state $|j\rangle$), and the Hamiltonian is invariant
and the dipole operator $\hat{P}$ is odd. Then odd harmonics alone
are generated because of the spatial-temporal symmetry. Other types
of spatial-temporal symmetry may lead to many interesting emission
patterns \cite{30sf2011NJP}.

Here we give some analysis on the hyper-Raman lines and their
relation to the harmonic components.  In our systems driven by a
periodic field, the quasienergy state has the form $|\psi_\alpha
(t)\rangle =e^{-i\varepsilon_\alpha t} |\phi_\alpha(t) \rangle$,
where the Floquet state $|\phi_\alpha(t) \rangle=|\phi_\alpha(t+T)
\rangle$  can be written in the form $|\phi_\alpha(t) \rangle=\sum
_m e^{-i m \omega_0 t} |\phi_\alpha^m \rangle$. We consider a state
$|\psi(t) \rangle = a_1 |\psi_1 \rangle+a_2 |\psi_2 \rangle=a_1
e^{-i\varepsilon_1 t}|\phi_1 \rangle+a_2 e^{-i \varepsilon_2
t}|\phi_2 \rangle$. Then we have
\begin{eqnarray}
&&\langle \psi | \hat{P} | \psi \rangle=
 \nonumber \\
 &&\sum_{m,n} e^{-i (n-m)\omega_0 t}
[|a_1|^2 \langle \phi_1^m | \hat{P} | \phi_1^n \rangle+|a_2|^2
\langle
\phi_2^m | \hat{P} | \phi_2^n \rangle ] \nonumber\\
&&+\sum_{m,n} e^{-i
[(\varepsilon_2-\varepsilon_1)+(n-m)\omega_0]t}a_2
a_1^* \langle \phi_1^m | \hat{P} | \phi_2^n \rangle  \nonumber\\
&&+\sum_{m,n}e^{-i
[(\varepsilon_1-\varepsilon_2)+(n-m)\omega_0]t}a_1 a_2^* \langle
\phi_2^m | \hat{P} | \phi_1^n \rangle .
\end{eqnarray}

 If the system posses a symmetry generated by the operator $Q=
 \Omega\cdot\theta$  and
 the Floquet state $|\phi_\alpha(t) \rangle$ has a definite parity
under Q, i.e., $Q|\phi_\alpha(t) \rangle=\pm |\phi_\alpha(t)
\rangle$, then we have $\Omega|\phi_\alpha^m \rangle=\pm (-1)^m
|\phi_\alpha^m \rangle$. Then we have  $\langle \phi_\alpha^m |
\hat{P} | \phi_\beta^n \rangle=0$ ($\alpha,\beta=1,2$) for $n-m$
even/odd number and states $\alpha$,$\beta$ of same/different
parity.  Quite often  the harmonics and the associated hyper-Raman
lines appear simultaneously due to the same symmetry properties of
the Floquet states. In the case with a monochromatic laser, we have
odd harmonics and the associated hyper-Raman lines. While in other
cases with symmetry broken, more harmonics and the associated
hyper-Raman lines are generated.

{\bf Emission patterns}
\begin{figure}
\begin{center}
\includegraphics*[angle=0, width= 0.48\textwidth]{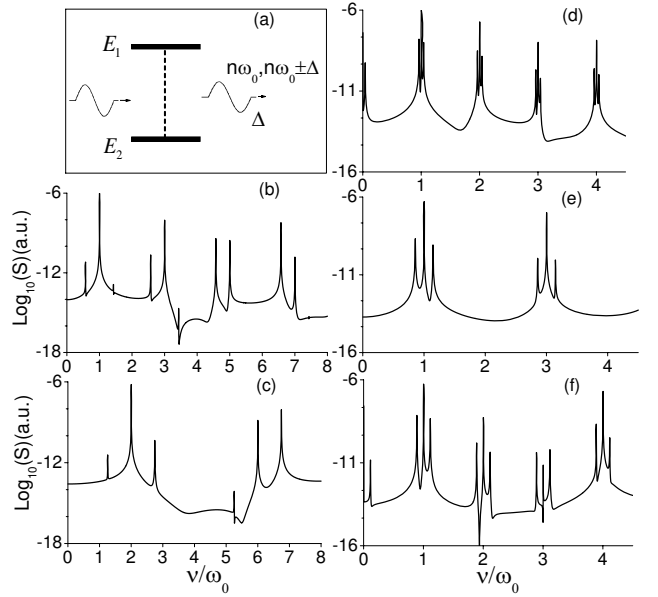}
\caption{ (a) The schematic diagram of our two-level system driven
by an incident laser. The emission spectrum contains the
higher-order harmonics and the accompanied hyper-Raman lines;
(b)-(d) Emission spectra in the presence of various driving fields
$F(t)$. (b) $F(t)=F_1 \cos(\omega_0 t)$; (c) $F(t)=F_2
\cos(2\omega_0 t)$; (d) $F(t)=F_1 \cos(\omega_0 t)+F_2
\cos(2\omega_0 t)$; (e) $F(t)=F_1 \cos(\omega_0 t)+F_2
\cos(3\omega_0 t)$; (f) $F(t)=F_1 \cos(\omega_0 t)+F_2
\cos(4\omega_0 t)$. $F_1=4.4 \times 10^{9} V/m, F_2=2.2 \times
10^{9} V/m$.} \label{FIG:SP}
\end{center}
\end{figure}
Let's first look at the emission spectrum of a system driven by a
monochromatic incident laser with frequency $\omega_0$. As seen in
Fig.1(b), there are odd harmonics as well as the associated
hyper-Raman lines due to the spatial-temporal symmetry  properties
of the Floquet states as analyzed above.  It is natural that, for
the incident laser with frequency $2\omega_0$, the emission spectrum
contains components with frequencies of $2\omega_0$, $6\omega_0$ and
$10\omega_0$...(odd orders of incident frequency $2\omega_0$) as
shown in Fig.1(c). One notices that there is no 0th hyper-Raman line
in this case. If we use two-color lasers with frequencies $\omega_0$
and $2\omega_0$, interesting phenomena appear. As seen from
Fig.1(d), even harmonics and the associated hyper-Raman lines are
generated by introducing the second laser field. Especially a low
frequency radiation, 0th hyper-Raman radiation is generated. We
would like to point out that the emission spectrum for the case with
driving field $F(t)=F_1 \cos(\omega_0 t)+F_2 \cos(2\omega_0 t)$ is
not the supposition of those driven by $F_1 \cos(\omega_0 t)$ and
$F_2 \cos(2\omega_0 t)$. For example, the harmonic $4\omega_0$ in
Fig. 1(d)  neither appears in Fig.1(b) nor in Fig.1(c). It is also
not the frequency summation or difference of the harmonics of
systems driven by  monochromatic incident laser with frequency
$\omega_0$ and $2\omega_0$. In fact, the appearance of all even
components (and the associated hyper-Raman lines) in Fig. 1(d) is
the consequence of symmetry breaking as discussed above.

According to our theory, the symmetry of the quantum system
generated by $Q_1$ is broken by introducing the second laser with
frequency $2k\omega_0$ ($k=1,2,3...$), and it is not broken by
introducing the second laser with frequency $(2k-1)\omega_0$
($k=1,2,3...$). These predictions are verified by our numerical
results shown in Fig.1(e) (the second laser of frequency
$3\omega_0$) and Fig.1(f) (the second laser of frequency
$4\omega_0$).
 It is clear that the 0th hyper-Raman line does not
appear in the cases with symmetry (see Figs.1(b) (c) (e)).
Interestingly, the second laser field with frequency $4\omega_0$
leads to the appearance of $2n$th harmonics (even for $n=1$) and the
associated hyper-Raman line (in particular the 0th hyper-Raman line)
as predicted by our theory, since the second laser with frequency
$4\omega_0$ breaks the spatial-temporal symmetry generated by $Q_1$.
Using this mechanism, we can explain the experimental results of
observing
even harmonics in helium or plasma plumes (containing nanoparticles,
carbon nanotubes, etc) driven by a two-color laser,\cite{29kim}
where the second-harmonic driving term, despite being very small,
breaks the symmetry and thus allows additional strong even harmonic
components. Our theory also naturally explains the generation of
even harmonics and associated hyper-Raman lines by breaking the
spatial symmetry \cite{6Kibis2009PRL}.

As seen from equation (3), the frequency of hyper-Raman line is
determined by the quasienergy, which is tunable. One may ask whether
one can use the hyper-Raman line associated with 1th harmonic to
generate the low frequency/THz wave? Actually, as seen from Fig.
2(a) the intensity of the hyper-Raman line associated with the 1th
harmonic decreases dramatically as the frequency decreasing. To make
this point clearer, we compare the low frequency components
associated the 1th harmonic and 0th harmonic for the case with
two-color laser shown in Fig. 2(b). One sees that in the every low
frequency regime, the intensity of the 0th hyper-Raman line
(associated with the 0th harmonic) is much larger than that of the
1th hyper-Raman line. Therefore one should use the 0th hyper-Raman
line
 to generate the THz radiation effectively.

\begin{figure}
\begin{center}
\includegraphics*[angle=0, width= 0.49\textwidth, height= 0.16\textheight]{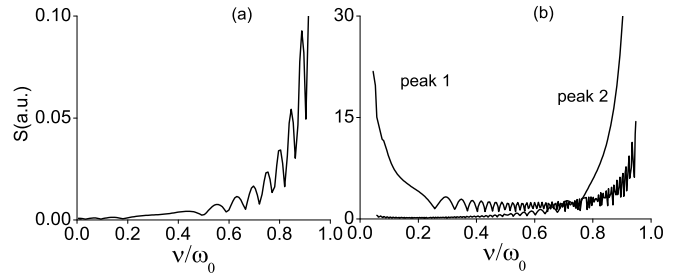}
\caption{(a) The intensity of hyper-Raman line associated with 1th
harmonic for $F(t)=F_1 \cos(\omega_0 t)$, $F_1$ is related to $\nu$,
the frequency of the hyper-Raman line; (b) The intensities of
hyper-Raman lines associated with 0th (peak 1) and 1th (peak 2)
harmonics for $F(t)=F_1 \cos(\omega_0 t)+F_2 \cos(2\omega_0 t)$,
$F_1=4.4 \times 10^{9} V/m$, $F_2$ is related to $\nu$.
}
 \label{FIG:01}
\end{center}
\end{figure}

{\bf Tuning of the frequency and intensity of THz wave}
\begin{figure}
\begin{center}
\includegraphics*[angle=0, width= 0.3\textwidth]{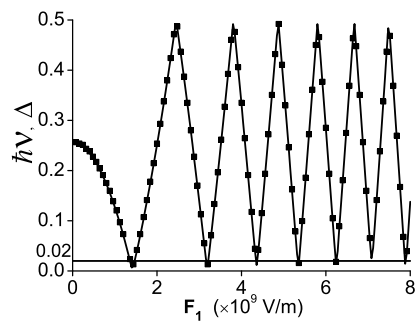}
\caption{The quasienergy difference (solid line) between two
quasi-eigenstates and the frequency (dots) of the 0th hyper-Raman
line versus the magnitude of external field $F_1$. $F_2=2.2 \times
10^{9} V/m$.} \label{FIG:QE}
\end{center}
\end{figure}
Now we discuss the optimization scheme of THz wave generation with
driving field $F(t)=F_1 \cos(\omega_0 t)+F_2 \cos(2\omega_0 t)$ in
our proposal. From equation (3), we see that  the difference of the
quasienergy is related  to the shift of the hyper-Raman line form
the corresponding harmonics, or the frequency of the 0th hyper-Raman
line. This relation is verified by our numerical calculation as
shown in Fig. 3. Moreover from Fig. 3, one
 sees that the 0th hyper-Raman line  with very small
frequency can be obtained in some parameter regimes, which can be
used for THz wave generation.  By applying this relation
$\Delta=|\epsilon_1-\varepsilon_2|=\hbar\nu$ ($\nu$ the frequency of
the 0th hyper-Raman line), any designated frequency THz wave can be
obtained by tuning the magnitudes of the two external fields $F_1$
and $F_2$. For example, THz radiation of 2THz ($0.02\omega_0$) can
be obtained by tuning the external field as shown in Fig. 3
[intersection points of the solid straight-line ($\nu=0.02\omega_0$)
and the line for quasienergy]. Fig. 4(a) shows the values for $F_1$
and $F_2$ under which the low frequency is $\nu=0.02\omega_0$. There
are disconnected parameter regimes of driving fields for generation
of THz radiation. Furthermore, the intensity of the THz wave can be
optimized. One can find the proper parameters for the maximal
intensity with desired frequency [the triangle in Fig. 4(a)].  The
corresponding emission spectrum is shown in Fig.4(b). It is seen
that the intensity of THz radiation has been increased 200 times
after optimization and we are able to obtain appreciably intense THz
wave (as strong as typical harmonics). Here we estimate the emission
power based on a system of arrays of semiconductor quantum quantum
dots. The emission power of each dot is $P\sim |\mu_{12}|^2
\nu^4/3c^3\sim 10^{-22} W$. There are about $N\sim 10^8$ dots in the
regime of size of wavelength
 which emit wave in phase. Therefore, the total
emission power from a sample of submillimeter size is around $N^2
P\sim  \mu W$. \cite{6Kibis2009PRL}

In our approach, the typical driving field intensity is in the order
of $10^{12}\sim 10^{13} W/cm^2$, which is lower than the typical
driving field intensity (in the order of $10^{14} W/cm^2 \sim
10^{15} W/cm^2)$ used in other methods for THz generation and/or HHG
by two-color laser \cite{fwmix,29kim,even-two}. One should also use
a long driving laser pulse [much longer than its period
($2\pi/\omega_0$)] as that used in Ref. [\onlinecite{29kim}]. The
emitted hyper-Raman line frequency can not only be tuned by the
driving field intensity as seen in Fig. 3, but also by the frequency
of the incident laser. If we use the incident laser with frequency
in the regime $\omega_0 \sim 50THz-500THz$, we may obtain the 0th
hyper-Raman line with frequency $\frac{1}{50} \omega_0 \sim
1THz-10THz$. In our method, there is another tunable parameter, the
phase difference between two incident lasers. The main physical
picture remains the same for different phases.

\begin{figure}
\begin{center}
\includegraphics*[angle=0, width= 0.48\textwidth, height= 0.16\textheight]{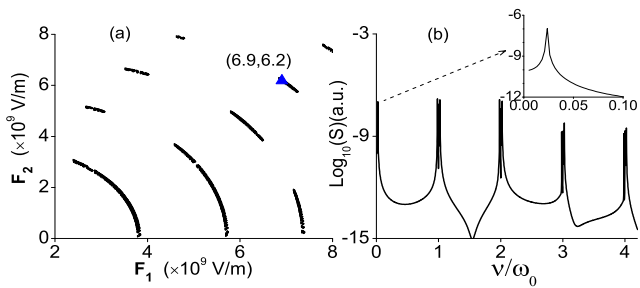}
 \caption{(a) The parameters of $F_1$ and $F_2$
 for the  0th hyper-Raman line with frequency 2THz$=0.02\omega_0$.
The triangle indicates the optimal parameter for
 the maximal intensity of 0th hyper-Raman line.
 (b) The corresponding emission spectrum for the case with optimized external
 field.
 } \label{FIG:OP}
\end{center}
\end{figure}

Unlike previous studies on non-linear driving two-level systems
\cite{two-nonlinear}, the hyper-Raman lines and low frequency
generation \cite{23Millack,25Zhou2008 ref24, 24Antonino,23a
Dakhnovskii ref21,HR}, our scheme of generating low frequency (THz)
wave is based on the symmetry principle in typical two-level quantum
systems of large energy spacing (in the order of $eV$). It can be
easily obtained from natural systems (such as atoms, molecules, and
semiconductors) and artificial structures (such as quantum dots).
The large energy spacing makes it robust against thermal fluctuation
and also insensitive to the initial state \cite{30sf2011NJP}.
Our mechanism based on the symmetry principle is different from
those (based on four-wave mixing \cite{fwmix} or transit current
\cite{current}) used before. Our non-perturbative analysis and
calculation show that one only needs to tune the intensity of the
two-color laser for typical two-level quantum systems, which is very
convenient in experiments. We may be able to obtain tunable THz
radiation due to the availability of stable, tunable driving sources
above the THz regime \cite{1Tonouchi2007,29kim}. Moreover, people
have already observed even harmonics in experiments with two-color
laser \cite{29kim,even-two}. Therefore it is quite likely that the
0th hyper-Raman line (in THz regime) associated with 0th harmonic
can be obtained in experiment if our optimization method is used.

{\bf Summary} A novel mechanism based on spatial-temporal symmetry
breaking is proposed to obtain THz wave from 0th hyper-Raman line in
two-color pumped two-level quantum system. Quasienergy is calculated
to determine the parameters of the incident laser to obtain
radiation of arbitrary frequency from 0.1THz to 10THz. Upon
optimization of the driving fields, we are able to obtain THz wave
with desired frequency and appreciable intensity (as large as that
of the typical harmonics).


{\bf Acknowledgments} This work was partially supported by the
National Science Foundation of China under Grants No.10874020,
11174042 and by the National Basic Research Program of China (973
Program) under Grants No. 2011CB922204, CAEP under Grant No.
2011B0102024, and the Project-sponsored by SRF for ROCS, SEM.



\begin{thebibliography}{}
\bibitem{1Tonouchi2007}
M. Tonouchi, nature photonics \textbf{1}, 97 (2007).
\bibitem{17Otsuji2006}
T. Otsuji, M. Hanabe, T. Nishimura, E. Sano, Opt. Express
\textbf{14}, 4815 (2006).
\bibitem{18Sekine2005}
N. Sekine, K. Hirakawa, Phys. Rev. Lett. \textbf{94}, 057408 (2005).
\bibitem{19Orihashi2005}
N. Orihashi, S. Suzuki, M. Asada, Appl. Phys. Lett. \textbf{87},
233501 (2005).
\bibitem{20Crowe2005}
T. W. Crowe, W. L. Bishop, D. W. Perterfi eld, J. L. Hesler, R. M.
Weikle, IEEE J. Solid-State Circuits \textbf{40}, 2104 (2005).

\bibitem{16Ito2005}
H. Ito, F. Nakajima, T. Furuta, T. Ishibashi, Semicond. Sci.
Technol. \textbf{20}, S191 (2005).
\bibitem{8Kawase2001}
K. Kawase, J. Shikata, I. Ito, J. Phys. D \textbf{34}, R1 (2001).


\bibitem{fwmix} X. Xie, J. Dai, and X.-C. Zhang, Phys. Rev. Lett. \textbf{96}, 075005
(2006); T.-J. Wang, J.-F. Daigle, S. Yuan, F. Theberge, M.
Chateauneuf, J. Dubois, G. Roy, H. Zeng, and S. L. Chin, Phys. Rev.
A \textbf{83}, 053801 (2011); T.-J. Wang, C. Marceau, Y. Chen, S.
Yuan, F. Theberge, M. Chateauneuf, J. Dubois, and S. L. Chin, Appl.
Phys. Lett. \textbf{96}, 211113 (2010); J. Penano, P. Sprangle, B.
Hafizi, D. Gordon, and P. Serafim, Phys. Rev. E \textbf{81}, 026407
(2010).

\bibitem{10Faist1994}
J. Faist, F. Capasso, D. L. Sivco, C. Sirtori, A. L. Hutchinson, and
A. Y. Cho, Science \textbf{264}, 553 (1994).
\bibitem{11Kohler2002}
 R. Kohler, A. Tredicucci, F. Beltram, H. E. Beere, E. H. Linfeld, A. G.
Davies, D. A. Ritchie, R. C. Iotti, and F. Rossi, Nature
\textbf{417}, 156 (2002).
\bibitem{14williams2007}
B. S. Williams, Nat. Photonics \textbf{1}, 517 (2007).
\bibitem{hu2004}B. S. Williams, S. Kumar, Q. Hu, and J. L. Reno, Electron. Lett. \textbf{40},
431 (2004).
\bibitem{9Belkin2007}
M. A. Belkin, F. Capasso, A. Belyanin, D. L. Sivco, A. Y. Cho, D. C.
Oakley, C. J. Vineis, and G. W. Turner, Nat. Photonics \textbf{1},
288 (2007).

\bibitem{21Williams2002}
 G. P. Williams, Rev. Sci.Instr. \textbf{73}, 1461 (2002).
\bibitem{22Bergner2005}
A. Bergner, U. Heugen, E. Brundermann, G. Schwaab, M. Havenith, D.
R. Chamberlin, and E. E. Haller, Rev. Sci. Instr. \textbf{76},
063110 (2005).

\bibitem{22a Ahn2007}
K. J. Ahn, F. Milde, and A. Knorr, Phys. Rev. Lett. \textbf{98},
027401 (2007).
\bibitem{22b Chassagneux2009}
Y. Chassagneux, R. Colombelli, W. Maineult, S. Barbieri, H. E.
Beere, D. A. Ritchie, S. P. Khanna, E. H. Linfield and A. G. Davies,
Nature(London) \textbf{457}, 174 (2009).
\bibitem{3zhang2009PRB}
S.Q. Duan, W. Zhang, Y. Xie, W.D. Chu, and X.G. Zhao, Phys. Rev. B
\textbf{80}, 161304(R) (2009).
\bibitem{23Millack}
T. Millack and A. Maquet, J. Mod. Opt. \textbf{40}, 2161 (1993).
\bibitem{30sf2011NJP}
 S. F. Guo, S. Q. Duan, Y. Xie, W. D. Chu and W. Zhang, New J. Phys. \textbf{13},
 053005 (2011).
\bibitem{30Narducci1990PRA}
L. M. Narducci, M. O. Scully, G.-L. Oppo, P. Ru, and J. R. Tredicce
Phys. Rev. A \textbf{42}, 1630 (1990).
\bibitem{29kim} R. A. Ganeev, H. Singhal, P. A. Naik, J. A. Chakera, H. S. Vora, R. A.
Khan and P. D. Gupta, Phys. Rev. A \textbf{82}, 053831 (2010); R. A.
Ganeev, H. Singhal , P. A. Naik, I. A. Kulagin, P. V. Redkin, J. A.
Chakera, M. Tayyab, R. A. Khan and P. D. Gupta, Phys. Rev. A
\textbf{80}, 033845 (2009).

\bibitem{6Kibis2009PRL}
O. V. Kibis, G. Ya. Slepyan, S. A. Maksimenko, and A. Hoffmann,
Phys. Rev. Lett. \textbf{102}, 023601 (2009).

\bibitem{even-two} I. J. Kim , C. M. Kim, H. T. Kim , G. H. Lee , Y. S. Lee, J. Y. Park, D. J. Cho
and C. H. Nam, Phys. Rev. Lett. \textbf{94}, 243901 (2005); I. J.
Kim, G. H. Lee, S. B. Park, Y. S. Lee, T. K. Kim, C. H. Namb, T.
Mocek and K. Jakubczak, Appl. Phys. Lett. \textbf{92}, 021125
(2008); N. Ishii, A. Kosuge, T. Hayashi, T. Kanai, J. Itatani, S.
Adachi, and S. Watanabe,Opt. express \textbf{16}, 20876 (2008).

\bibitem{two-nonlinear}K. B. Nordstrom, K. Johnsen, S. J. Allen, A.-P. Jauho, B. Birnir,
J. Kono,  T. Noda, H. Akiyama, and H. Sakaki, Phys. Rev. Lett.
\textbf{81}, 457 (1998); L. Plaja and L. Roso, J. Mod. Opt.
\textbf{40}, 793 (1993).

\bibitem{24Antonino}
A. D. Piazza and E. Fiordilino, Phys. Rev. A  \textbf{64}, 013802
(2001).
\bibitem{25Zhou2008 ref24}
Z. Y. Zhou and J. M. Yuan, Phys. Rev. A \textbf{77}, 063411 (2008).
\bibitem{23a Dakhnovskii ref21}
Y. Dakhnovskii and H. Metiu, Phys. Rev. A \textbf{48}, 2342 (1993).
\bibitem{HR} C. Liu, S. Gong, R. Li, and Z. Xu,
Phys. Rev. A \textbf{69}, 023406 (2004);  F. I. Gauthey, C. H.
Keitel, P. L. Knight, and A. Maquet, Phys. Rev. A \textbf{52}, 525
(1995); M. Frasca, Phys. Rev. A \textbf{60}, 573 (1995);  A. D.
Piazza, E. Fiordilino and M. H. Mittleman, Phys. Rev. A \textbf{64},
013414 (1995);  M. L. Pons, R. Taieb, and A. Maquet, Phys. Rev. A
\textbf{54}, 3634 (1996); H. Wang and X.-G. Zhao, J. Phys.: Condens.
Matter \textbf{8}, L285 (1996).
\bibitem{current} K. Y. Kim, J. H. Glownia, A. J. Taylor, and G. Rodriguez, Opt. Express \textbf{15}, 4577
(2007); K. Y. Kim, A. J. Taylor, J. H. Glownia, and G. Rodriguez,
Nat. Photon. \textbf{2}, 605 (2008); Y. Minami, M. Nakajima, and T.
Suemoto, Phys. Rev. A \textbf{83}, 023828 (2011).










\end{thebibliography}
\end{document}